\documentclass{iaus}
\usepackage{graphicx}

\title[Periodic variations in 6.7-GHz methanol masers] %% give here short title %%
{Periodic variations in 6.7-GHz methanol masers}

\author[Goedhart, Gaylard \& van der Walt]   %% give here short author list %%
{Sharmila Goedhart$^1$, Michael J. Gaylard$^1$, D. Johan van der Walt$^2$}

\affiliation{$^1$Hartebeesthoek Radio Astronomy Observatory, PO Box 443, Krugersdorp 1740, South Africa\break email: sharmila@hartrao.ac.za\\[\affilskip]
$^2$ Unit for Space Physics, North-west University, South Africa }

\pubyear{2007}
\volume{242}  %% insert here IAU Symposium No.
\pagerange{}
\date{?? and in revised form ??}
\jname{Astrophysical masers and their environments}
\editors{J. Chapman \& W. Baan, eds.}

\begin{document}

\maketitle

\begin{abstract}
An intensive monitoring program of 54 6.7-GHz methanol maser sources was carried out at the Hartebeesthoek Radio Astronomy Observatory from January 1999 to April 2003.  The monitoring program was subsequently continued on 19 sources of interest.  Analysis of the resulting time-series stretching over eight years shows that six of the sources are periodic, with periods ranging from 133 days to 504 days.  The waveforms in individual sources range from sinusoidal fluctuations to sharp flares and there can be other long term trends in the time-series. The amplitudes of the variations can also change from cycle to cycle.  The time-series of the periodic masers will be presented, and possible causes of the variability discussed.
\end{abstract}

\section{Introduction}

The existence of variability in 6.7-GHz masers was established by \cite{Cas95c} and \cite{Mac96}. However, the nature of the variability had not been fully characterised until the study of \cite{Goe04} who monitored 54 sources for four years. The results of this study showed a range of variability behaviour, including monotonic, quasi-periodic and  what appeared to be periodic variations. The monitoring programme has been continued on the most variable sources, enabling the confirmation of periodicity in six sources, despite considerable upgrades to the telescope during this time.  The time-series of the periodic candidates are presented here.

\section{Observations}

The observations were done using the 26-m telescope at the Hartebeesthoek Radio Astronomy Observatory.  The observations from January 1999 to April 2003 were of left-circular polarisation using a 256-channel spectrometer. Observations resumed in September 2003 with the cryogenic receiver upgraded from single to dual polarisation, a 2x1024 channel spectrometer and a new telescope control system. The telescope surface was replaced by solid panels, with the final alignment taking place in September 2004.

Amplitude calibrations were based on monitoring of Virgo A, Hydra A and 3C123.  The calibrations were checked against the relatively quiescent source G351.42+0.64.

\section{Results}

Periodicity analysis was done on time-series extracted from velocity channels at the maser peaks.  Fourier analysis was done using the program \textsc{Period04} developed by \cite{Len05}. Since many of the time-series are non-sinusoidal, the periods were confirmed by epoch-folding using the test statistic developed by \cite{Dav90}. 

\begin{figure}
 \resizebox{\hsize}{!}{\includegraphics[clip]{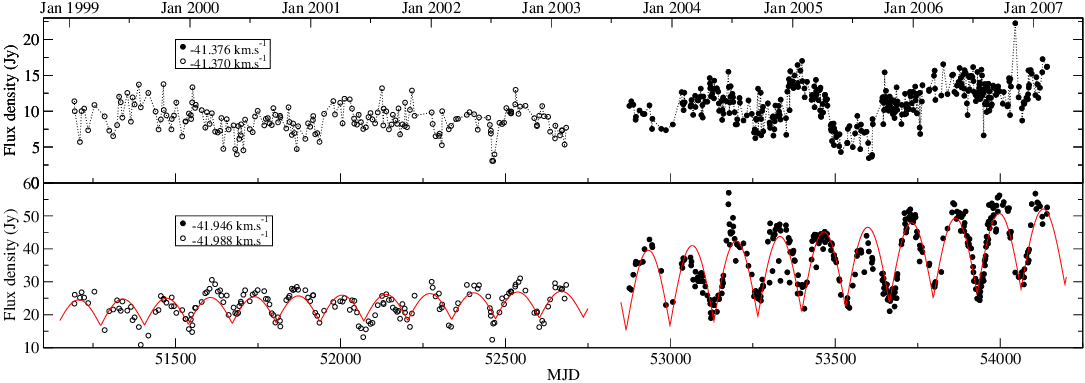}}
  \caption{Time-series for selected peak velocity channels in G338.93-0.06. The solid curves show the result of fitting a first-order polynomial plus $\vert{\sin}\vert$.}\label{fig:g3389-ts}
\end{figure}

Time-series at velocity channels of interest will be presented. Figure~\ref{fig:g3389-ts} shows the time-series for the two strongest spectral features in G338.93-0.06. The intensity variations in a single velocity channel are shown. The variations in the two features are uncorrelated, with the feature at -41.9 km s$^{-1}$ showing regular variations with a period of 133 days.  Nineteen cycles have been observed to date in this source, with the overall intensity of the periodic feature steadily increasing. The amplitude of the variations have also been increasing. The shape of the light-curve can be fitted by a function of the form $\vert{\sin}\vert$.

\begin{figure}
 \resizebox{\hsize}{!}{\includegraphics[clip]{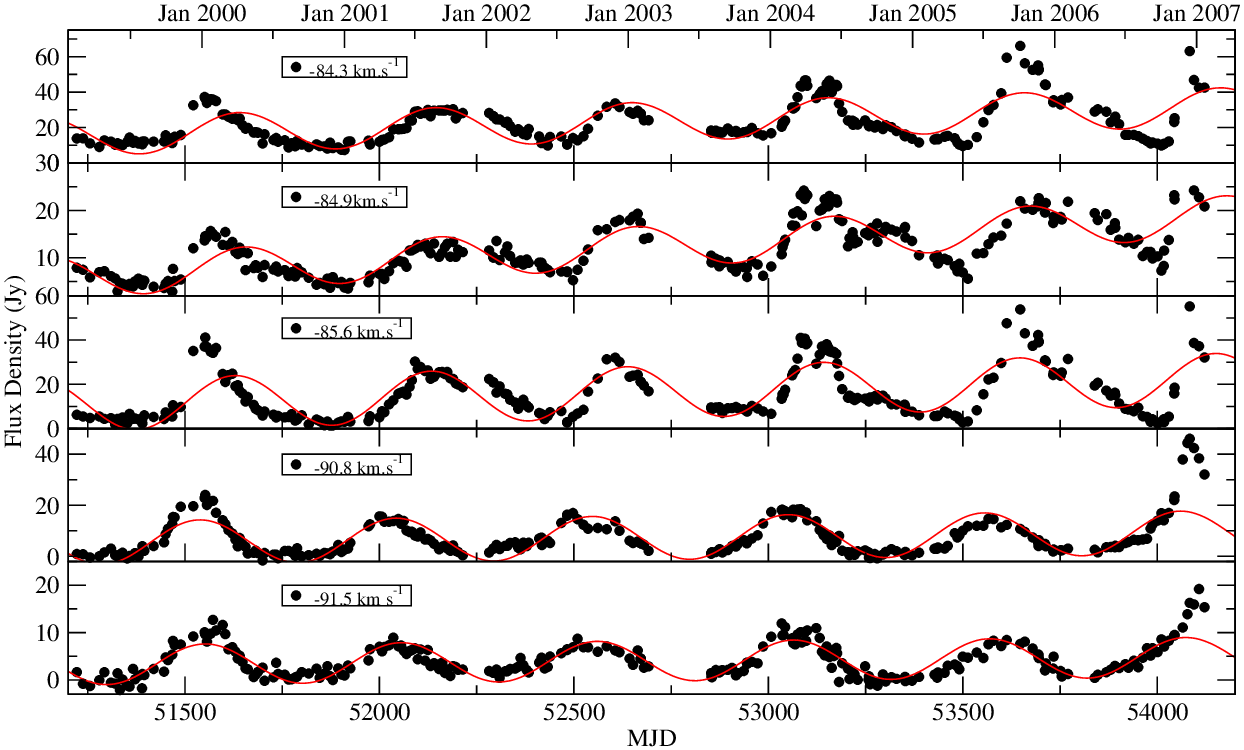}}
  \caption{Time-series for selected peak velocity channels in G331.13-0.24. The solid curves show the results of fitting a sinusoid plus a first order polynomial.}\label{fig:g3311-ts}
\end{figure}

G331.13-0.24 shows variations with a period of 504 days (Figure~\ref{fig:g3311-ts}). The maser features in the velocity range -92 to -89 km s$^{-1}$ show extremely regular variations which follow a sinusoidal form for part of the cycle. The group at -87 to -83.5 km s$^{-1}$ show regular variations with a time-delay from the start  of the flares at the first group. The durations of these flares vary, as well as the shape of the flares but the start times of the flares appear to be periodic. 

\begin{figure}
 \resizebox{\hsize}{!}{\includegraphics[clip]{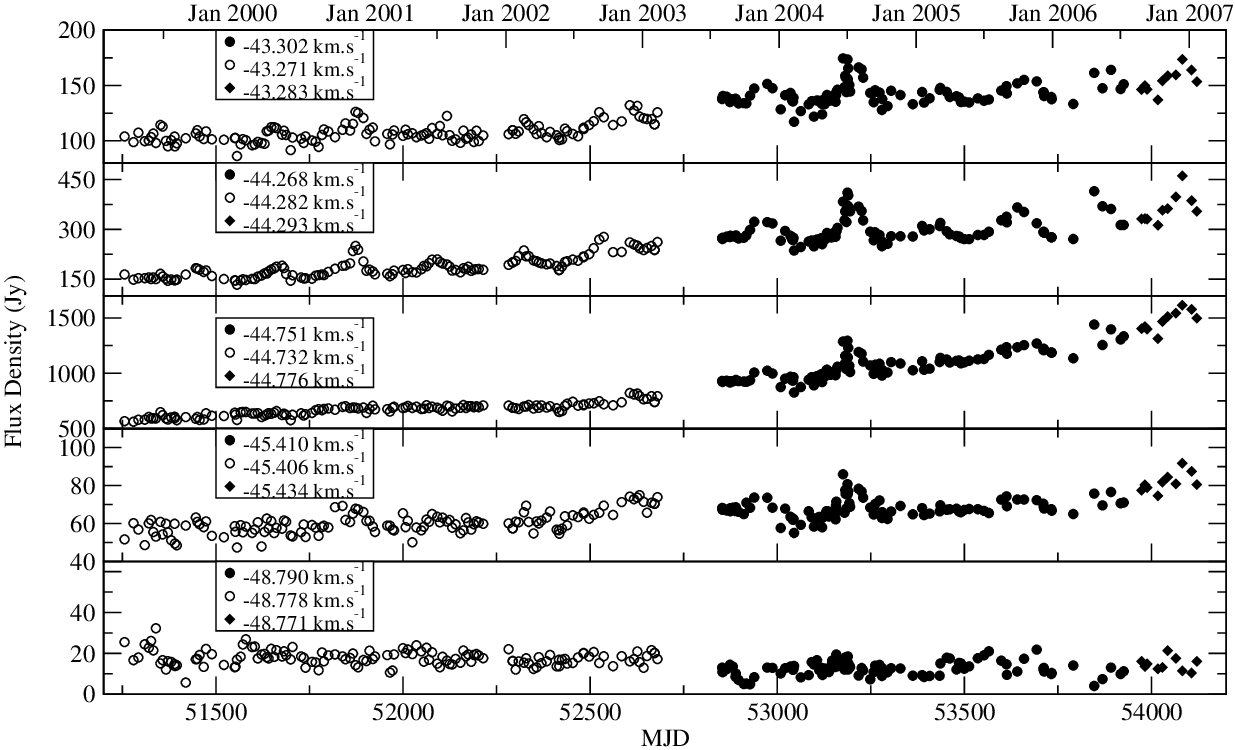}}
  \caption{Time-series for selected peak velocity channels in G328.24-0.55.}\label{fig:g3282-ts}
\end{figure}

G328.24-0.55 shows clear periodicity with a period of 220 days in only one spectral feature (Figure~\ref{fig:g3282-ts}) at -44.3 km s$^{-1}$. Weak fluctuations with the same period are seen in the feature at -43.3  km s$^{-1}$. Interestingly, a large flare in June 2004 was seen at all features in the velocity range -46 to -43 km s$^{-1}$.

\begin{figure}
 \resizebox{\hsize}{!}{\includegraphics[clip]{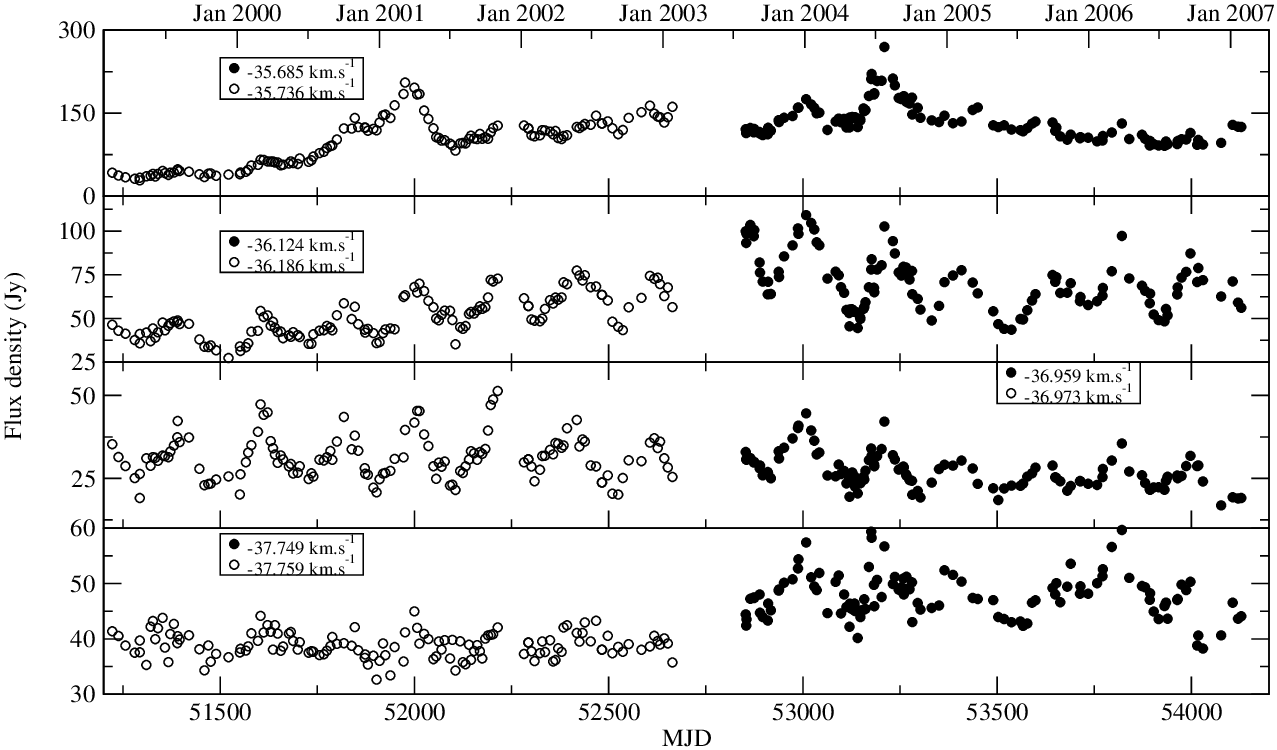}}
  \caption{Time-series for selected peak velocity channels in G339.62-0.12.}\label{fig:g3396-ts}
\end{figure}

G339.62-0.12 shows periodic variations with a period of 201 days in all the spectral features (Figure~\ref{fig:g3396-ts}). The peak at -35.7 km s$^{-1}$ generally has low amplitude variations but  occasionally shows very strong flares.

G9.62+0.20 shows periodic flaring with a period of 244 days.  This source is known to show simultaneous flaring at 12.2 GHz with an even greater amplitude (\cite{Goe03}; Gaylard \& Goedhart, this volume).

\begin{figure}
 \resizebox{\hsize}{!}{\includegraphics[clip]{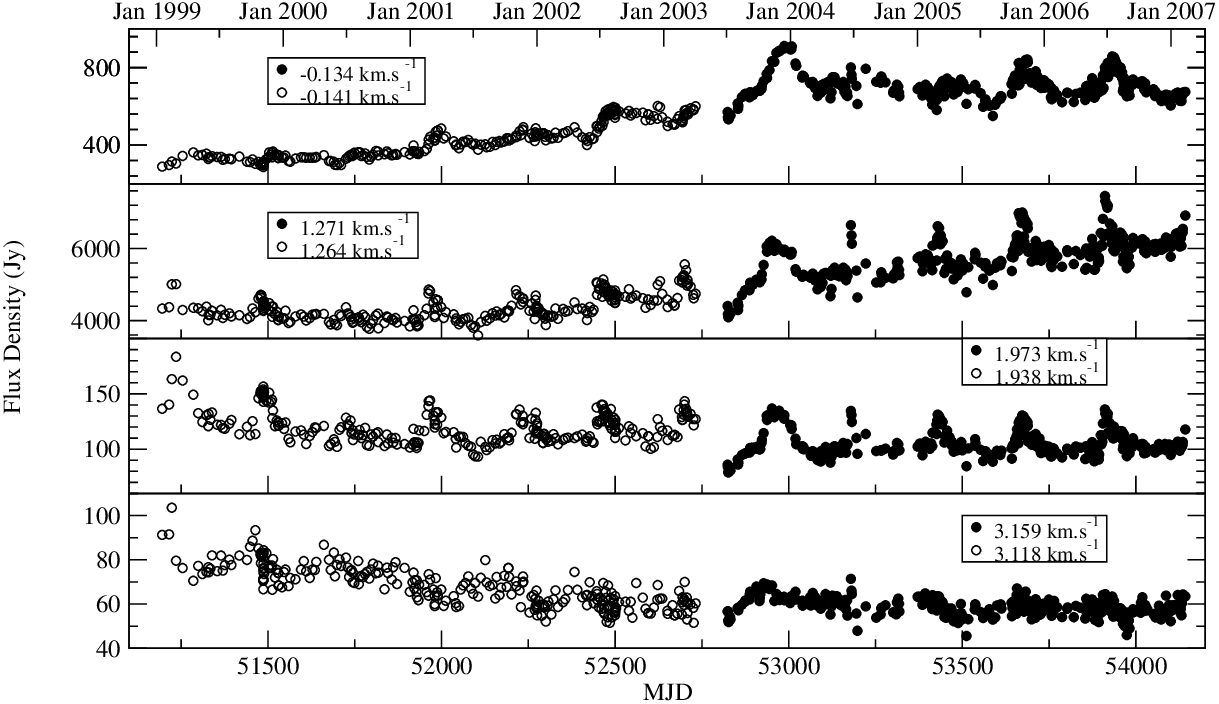}}
  \caption{Time-series for selected peak velocity channels in G9.62+0.20.}\label{fig:g0096-ts}
\end{figure}

\begin{figure}
 \resizebox{\hsize}{!}{\includegraphics[clip]{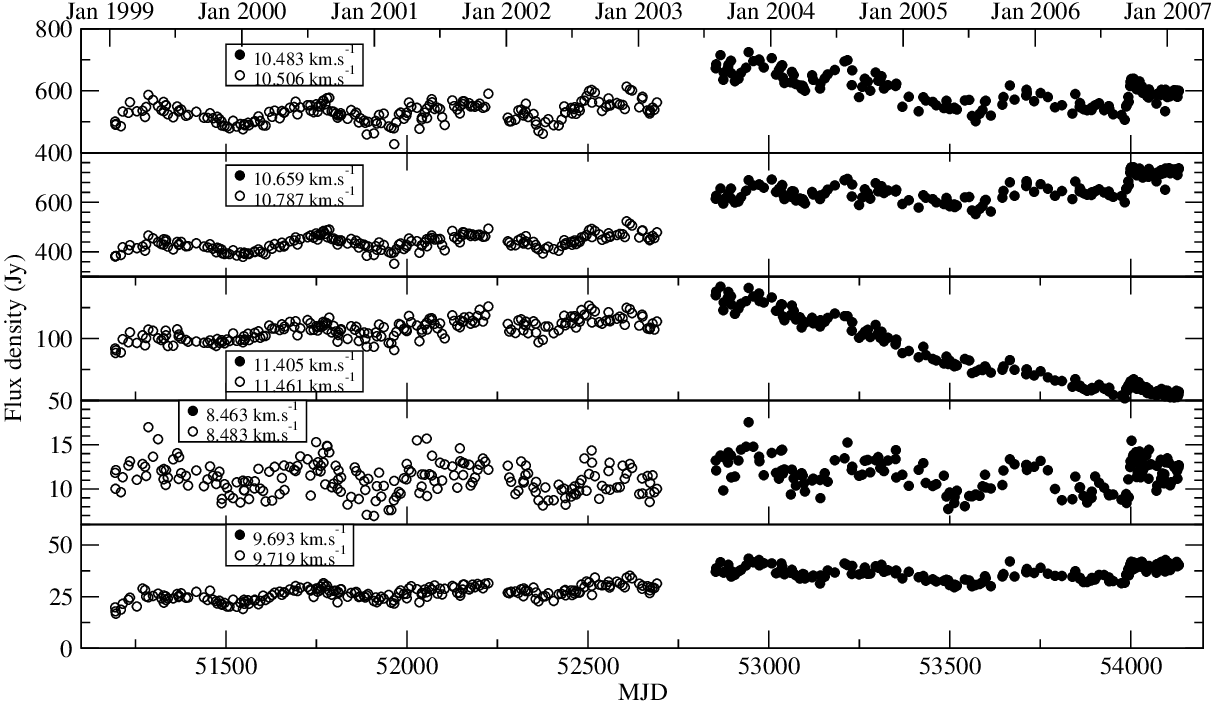}}
  \caption{Time-series for the peak velocity channels in G188.95+0.89.}\label{fig:g1889-ts}
\end{figure}

G188.95+0.89 showed low amplitude sinusoidal variations with a period of 404 days (Figure~\ref{fig:g1889-ts}). However, the behaviour of this source has been changing since mid-2003, with the spectral feature at -11.4 km s$^{-1}$ decreasing steadily in intensity. Linear detrending of the time-series reveals that the sinusoidal variations persisted until $\sim$mid-2006.  There was a sharp increase in intensity in August 2006, marking the time at which the time-series no longer show sinusoidal behaviour.  This source may be at the end of the periodic phase in its lifetime or the cycle pattern is changing.  Further monitoring will be necessary to see what happens next.

\section{Discussion}

The environment in which methanol masers are found is complex, with a number of factors that could affect the maser intensity.  The radiation for pumping the masers originates at the central star but will have been considerably reprocessed by the time it arrives at the maser region. The central (proto)star may be surrounded by a rotating, inhomogeneous disk which may also be feeding an outflow. The outflows may be sporadic or there may be precessing jets. The central object may even be a close binary which could still have associated accretion disks, leading to complex dynamical interactions.  The source of the pump photons, believed to be the HII region, could also be variable. Interpretation of the variability is also limited by the uncertainty of where the masers are situated in the circumstellar environment. Current observational evidence is unclear but the masers may be tracing circumstellar disks, outflow cavities or weak shocks.  Periodic mechanisms can be invoked for any of these scenarios.
  
The variations in the masers shown here appear to have a periodic component with periods in the range 133 to 504 days. This must be pointing to a specific mechanism. The typical periods are much larger than those seen in pulsating stars, with the exception of long-period variables such as Miras.  However, these are evolved stars and unlikely to be found in the vicinity of the methanol masers.  O and B type stars have extremely rapid rotation rates, around 0.5 to 2 days, which are far too short to be related to the observed periodicity. Keplerian orbits within 1 to 10 AU of the central star have periods in the observed range.  However, it is not clear what the orbiting objects would be or how this would lead to a modulation in the masers. Since there is a high incidence of binarity or multiplicity in optically visible stars, it is not inconceivable that a binary system could be present. On the other hand, we may be seeing a process related to an accretion disk.

\section{Conclusion}

Periodicity was found in $\sim$ 10\% of the original sample of 54 sources monitored. No clear cause of the variations can be found at this stage.  High-resolution, multi-wavelength observations of the methanol maser sources are necessary to understand conditions in these regions.  Detailed modelling will also be necessary to understand the effects of radiative transfer in different morphologies and maser models with time-dependent components would have to be developed. This may be a way to gain greater insight into processes occurring in regions that can not be directly observed.

\end{document}